\documentclass[twocolumn]{svjour3}          
\smartqed  
\usepackage{graphicx}
\usepackage{bm}        
\usepackage{amssymb}   
\usepackage{cite}
\usepackage{color}
\begin{document}
\title{From strangelets to strange stars: A unified description}


\author{ Cheng-Jun Xia  \and Guang-Xiong Peng   \and   En-Guang Zhao  \and Shan-Gui Zhou}


\institute{C.-J. Xia$^{1}$,
              \email{cjxia@itp.ac.cn}        \and
              G.-X. Peng$^{2,3,4}$ \and
              E.-G. Zhao$^{1,5}$   \and
              S.-G. Zhou$^{1,4,5}$
\at$^{1}${State Key Laboratory of Theoretical Physics, Institute of Theoretical Physics, Chinese Academy of Sciences, Beijing 100190, China}
\\$^{2}${School of Physics, University of Chinese Academy of Sciences, Beijing 100049, China}
\\$^{3}${Theoretical Physics Center for Science Facilities, Institute of High Energy Physics, Beijing 100049, China}
\\$^{4}${Synergetic Innovation Center for Quantum Effects and Application, Hunan Normal University, Changsha 410081, China}
\\$^{5}${Center of Theoretical Nuclear Physics, National Laboratory of Heavy Ion Accelerator, Lanzhou 730000, China}}


\maketitle

\begin{abstract}
The conventionally separated treatments for strangelets and strange stars are now unified with a more comprehensive theoretical
description for objects ranging from strangelets to strange stars. After constraining the model parameter according to the
Witten-Bodmer hypothesis and observational mass-radius probability distribution of pulsars, we investigate the properties of
this kind of objects. It is found that the energy per baryon decreases monotonically with increasing baryon number and reaches
its minimum at the maximum baryon number, corresponding to the most massive strange star. Due to the quark depletion, an electric
potential well is formed on the surface of the quark part. For a rotational bare strange star, a magnetic field with the typical
strength in pulsars is generated.
\end{abstract}

\vspace{1cm}

\section{\label{sec:intro}Introduction}

It was pointed out long ago that strange quark matter (SQM) might be the ground state of strongly interacting matter, which is nowadays
called the Witten-Bodmer hypothesis~\cite{Bodmer1971_PRD4-1601, Witten1984_PRD30-272}. If true, there should exist stable objects of SQM
with the baryon number $A$ ranging from a few to $\sim$$10^{57}$. Customarily, small SQM nuggets with $A\lesssim 10^7$ are often
referred to as strangelets~\cite{Berger1987_PRC35-213, Madsen1993_PRL70-391, Farhi1984_PRD30-2379, Greiner1987_PRL58-1825,
Gilson1993_PRL71-332, He1996_PRC53-1903, Wen2007_JPG34-1697,  Xia2014_SCPMA57-1304}, or slets~\cite{Peng2006_PLB633-314},
while stars consisting of SQM are called strange (quark) stars~\cite{Alcock1986_ApJ310-261, Weber2005_PPNP54-193, Itoh1970_PTP44-291,
Haensel1986_AA160-121, Perez-Garcia2010_PRL105-141101, Herzog2011_PRD84-083002, Dexheimer2013_EPJC73-2569, Chang2013_SCPMA56-1730,
Xia2014_PRD89-105027, Xu2015_PRD92-025025}, being possible candidates for pulsars.

Lumps of SQM are expected to be produced in the collision of binary compact stars containing SQM~\cite{Madsen2005_PRD71-014026,
Bauswein2009_PRL103-011101}. Further collisions among those lumps may create slets, nuclearites~\cite{Rujula1984_Nature312-734,
Lowder1991_NPB24-177}, meteorlike compact ultradense objects~\cite{Rafelski2013_PRL110-111102} etc., and some of them may eventually make
their way to our Earth~\cite{Monreal2007_JHEP02-077}. Due to the special characteristics of these objects such as the lower charge-to-mass
ratio~\cite{Sandweiss2004_JPG30-S51, Han2009_PRL103-092302}, the larger mass~\cite{Herrin2006_PRD73-043511}, the highly ionizing
tracks in the interstellar hydrogen cloud (e.g., pulsar scintillations)~\cite{Pcuteerez-Garccuteia2013_PLB727-357}, and the characteristic
gamma rays through heavy ion activation~\cite{Isaac1998_PRL81-2416}, there are possibilities to observe them. However, despite decades
of efforts, no compelling evidence for the existence of stable SQM is found  (for reviews, see, e.g., Refs.~\cite{Klingenberg1999_JPG25-R273,
Finch2006_JPG32-S251}).

This case is due to the extreme complexity of an SQM system which involves all the fundamental interactions, i.e., the strong, weak,
electromagnetic, and gravitational interactions. In the conventional theoretical treatments, significantly
different simplifications were adopted for slets and strange stars. For a slet, electrons were
ignored since the Compton wavelength is much larger than the size of the quark part~\cite{Madsen1999_LNP516-162}, and quarks were
assumed to be uniformly distributed. For strange stars, gravity has to be considered. The normal way is to first get an equation
of state of SQM by assuming the local charge neutrality, and then obtain the mass-radius ($M$-$R$) relation by solving the
Tolman-Oppenheimer-Volkov equations.

However, according to recent studies, effects such as the charge screening, electron-positron pair creation, and nonzero charge
densities in strange stars have important implications on the properties of SQM. For example, taking into account the electrostatic
effects, Alford et al.~\cite{Alford2006_PRD73-114016} found that, for a small enough surface tension, large slets are
unstable to fragmentation and strange star surfaces fragment into a crystalline crust made of slets and electrons.
For quark-hadron phase transition, the finite-size effect turns out to be very important~\cite{Voskresensky2002_PLB541-93,
Tatsumi2003_NPA718-359, Voskresensky2003_NPA723-291, Endo2005_NPA749-333}. It was shown that the geometrical structures may be
destabilized by the charge screening effect~\cite{Maruyama2007_PRD76-123015}. Due to the electron-positron pair creation, an upper
bound on the net charge of slets or strange stars was found~\cite{Madsen2008_PRL100-151102}. The local charge neutrality in
compact stars is also in question~\cite{Rotondo2011_PLB701-667}. In the case of a neutron star, an overcritical electric field was
found in the transitional region from the core to the crust~\cite{Belvedere2012_NPA883-1}. For a bare strange star, an electric dipole
layer may be formed on the surface and results in an electric field of $\sim$$10^{17-19}\ \mathrm{V}/\mathrm{cm}$~\cite{Alcock1986_ApJ310-261}.
Due to the presence of a critical electric field, the electron-positron production may be induced and results in some astrophysical
observables~\cite{Ruffini2010_PR487-1}. The mass and radius of a strange star are increased by $\sim$$15\%$ and $\sim$$5\%$, respectively,
if the star possesses a net charge on the surface~\cite{Negreiros2009_PRD80-083006}.

Meanwhile, the possibility of pulsars being strange stars may give us an insight into the properties of SQM. Up till now, around 2,500
pulsars have been observed and among them about 70 pulsars'  masses were measured~\cite{Manchester2005_AJ129-1993, Manchester2015}. At the
same time, more than 10 pulsars provide us the $M$-$R$ probability distributions with photospheric radius expansion bursts as well as
quiescent low-mass X-ray binaries~\cite{Lattimer2012_ARNPS62-485, Steiner2010_ApJ722-33, Guillot2013_ApJ772-7, Lattimer2014_EPJA50-40,
Li2015_ApJ798-56}. If SQM is absolutely stable, those pulsars may be strange stars~\cite{Page2006_ARNPS56-327}, then the properties of
SQM can be constrained with the $M$-$R$ relations.

In the present paper,  we study the SQM system ranging from slets to strange stars in a unified description.
After constraining the only model parameter, the bag constant $B$, according to the Witten-Bodmer hypothesis and the observational
$M$-$R$ probability distribution of pulsars, it is found that the ratio of charge to baryon number of a slet is different
from previous findings, while the size is significantly smaller than that of a nucleus with the same mass number. In addition,
rotation of a bare strange star generates a strong magnetic field with the typical strength in pulsars.
This paper is organized as follows.  In Section~\ref{sec:the}, the unified description for SQM objects is presented, where
the effects of gravity and electrostatic interactions are treated on the macroscopic scale while the strong and weak interactions
are considered locally. Based on this description, the properties of SQM objects ranging from slets to strange stars are investigated
in Section~\ref{sec:rst}. A summary is presented in Section~\ref{sec:sum}.

\section{\label{sec:the}Theoretical framework}
The internal structure of a spherically symmetric, charged, and static object should fulfill the thermodynamic equilibrium condition,
which can be obtained by minimizing the energy of the system for given total particle number and entropy. We consider the gravity and
electrostatic interactions on the macroscopic scale. The metric for the SQM sphere reads
\begin{equation}
  \mbox{d}s^2=\mathrm{e}^\nu \mbox{d}t^2 - \mathrm{e}^\lambda \mbox{d}r^2 - r^2 (\mbox{d}\theta^2 +\sin^2{\theta}\mbox{d}\phi^2), \label{eq:metric}
\end{equation}
where $r$, $\theta$, and $\phi$ are the standard spherical coordinates with the metric elements satisfying
\begin{eqnarray}
\mathrm{e}^{-\lambda} &=& 1- \frac{2G}{r}M_\mathrm{t},
\label{eq:elmda} \\
\frac{\mbox{d}\nu}{\mbox{d}r} &=& \frac{2 G \mathrm{e}^{\lambda}}{r^2}
                                  \left[4\pi  r^3  \left( P - \frac{\alpha Q^2}{8\pi r^4} \right) +  M_\mathrm{t} \right].
\label{eq:dnu}
\end{eqnarray}
Here we use the natural system of units, with $G$ and $\alpha$ being the gravitational and fine-structure constants. The total
mass, particle number, and entropy are obtained with
\begin{eqnarray}
M_\mathrm{t}(r) &=& \int_0^r 4\pi r'^2 \left( E+{\alpha Q^2}/{8\pi r'^4}\right)\mbox{d}r',
\label{eq:Mt} \\
N_i(r)  &=& \int_0^r 4\pi  n_i(r') \mathrm{e}^{\lambda/2}r'^2 \mbox{d}r',
\label{eq:Ni} \\
\bar{S}(r) &=& \int_0^r 4\pi  S(r') \mathrm{e}^{\lambda/2}r'^2 \mbox{d}r'.
\label{eq:S}
\end{eqnarray}
Then the total charge is given by $Q(r) = \sum_i q_i N_i(r)$ with $q_u=2/3$, $q_d=q_s=-1/3$, and $q_e=-1$.
Based on the Thomas-Fermi approximation, the pressure $P(r)$, energy density $E(r)$, particle number
density $n_i(r)$, and entropy density $S(r)$ are given locally by incorporating both the strong and weak interactions.

By minimizing the mass $M=M_\mathrm{t}(\infty)$ with respect to the particle distribution $N_i(r)$ and entropy
distribution $\bar{S}(r)$ at the fixed total particle number $N_i(\infty)$ and entropy $\bar{S}(\infty)$,
we immediately have
\begin{eqnarray}
\frac{\mbox{d}\mu_i}{\mbox{d}r} &=& \frac{Q}{r^2}q_i\alpha \mathrm{e}^{\lambda/2} - \frac{\mu_i}{2} \frac{\mbox{d}\nu}{\mbox{d}r}, \label{eq:pdis}\\
   \frac{\mbox{d}T}{\mbox{d}r}  &=& - \frac{T}{2} \frac{\mbox{d}\nu}{\mbox{d}r},  \label{eq:sdis}
\end{eqnarray}
with $\mu_i(r)$ and $T(r)$ being the chemical potential and temperature.

For the local properties of SQM, we adopt the bag model and consider only zero temperature, where the thermodynamic potential
density is given by
\begin{equation}
 \Omega(r)=\Omega_0(\{\mu_i(r)\}, \{m_i\}) + B
\end{equation}
in the ideal Fermi-gas approximation. To reach the lowest energy, SQM undergoes weak reactions and reach the chemical equilibrium
\begin{equation}
  \mu_u + \mu_e=\mu_d=\mu_s.
\end{equation}
Then the internal structure of
an SQM sphere can be determined by solving Eq.~(\ref{eq:pdis}). Since electrons are not confined by the vacuum pressure, an atom-like
structure of the SQM system is formed, i.e., a positively charged SQM core with a cloud of electrons surrounding it.

The quark-vacuum interface on the surface of the SQM core needs to be treated with special care. We consider the number of depleted
quarks on the interface by adopting the multiple reflection expansion (MRE) method~\cite{Berger1987_PRC35-213, Madsen1993_PRL70-391}
\begin{equation}
\frac{\mbox{d} N_i^\mathrm{surf}}{\mbox{d} p_i}
 = \frac{2 g_i R}{3\pi} -\frac{g_i p_i R}{m_i \pi} (m_i R+1) \mathrm{arctan}\left(\frac{m_i}{p_i}\right).
\label{eq:state_surf}
\end{equation}
Here $p_i$ is the momentum of quark flavor $i$ ($i=u,d,s$). Its upper bound corresponds to the Fermi momentum on the surface $\nu_i(R)$ with $R$ being
the radius of the SQM core. Note that Eq.~(\ref{eq:state_surf}) only gives the average number of depleted quarks, while for
smaller systems shell corrections may be important~\cite{Madsen1994_PRD50-3328}.
Then the energy contribution and pressure are given by
\begin{equation}
 \bar{E}_i^\mathrm{surf} = \int_0^{\nu_i(R)}\sqrt{p_i^2+m_i^2} \frac{\mbox{d} N_i^\mathrm{surf}}{\mbox{d} p_i} dp_i
\end{equation}
and
\begin{equation}
 P^\mathrm{surf} =-\sum_{i} \left.\frac{\mbox{d} \bar{E}_i^\mathrm{surf} }{\mbox{d} V}\right|_{N_i^\mathrm{surf}}.
\end{equation}
Under the influence of gravity,
the energy contribution to the mass is $M^\mathrm{surf}=\sum_i\bar{E}_i^\mathrm{surf}\mathrm{e}^{\nu(R)/2}$.

The quark-vacuum interface is obtained when the pressure of quarks is in balance with the vacuum pressure, i.e.,
\begin{equation}
P(R) - P_e(R) + P^\mathrm{surf} = 0. \label{eq:P_stable}
\end{equation}
Then we have the total quark number $N_q = N_q(R) + N_q^\mathrm{surf}$, mass $M=M_\mathrm{t}(\infty)+M^\mathrm{surf}$,
and charge of the core $Q(R) = \sum_i q_i\left[N_i(R) + N_i^\mathrm{surf}\right]$.

\section{\label{sec:rst}Results and discussions}

\begin{figure}
\includegraphics[width=\linewidth]{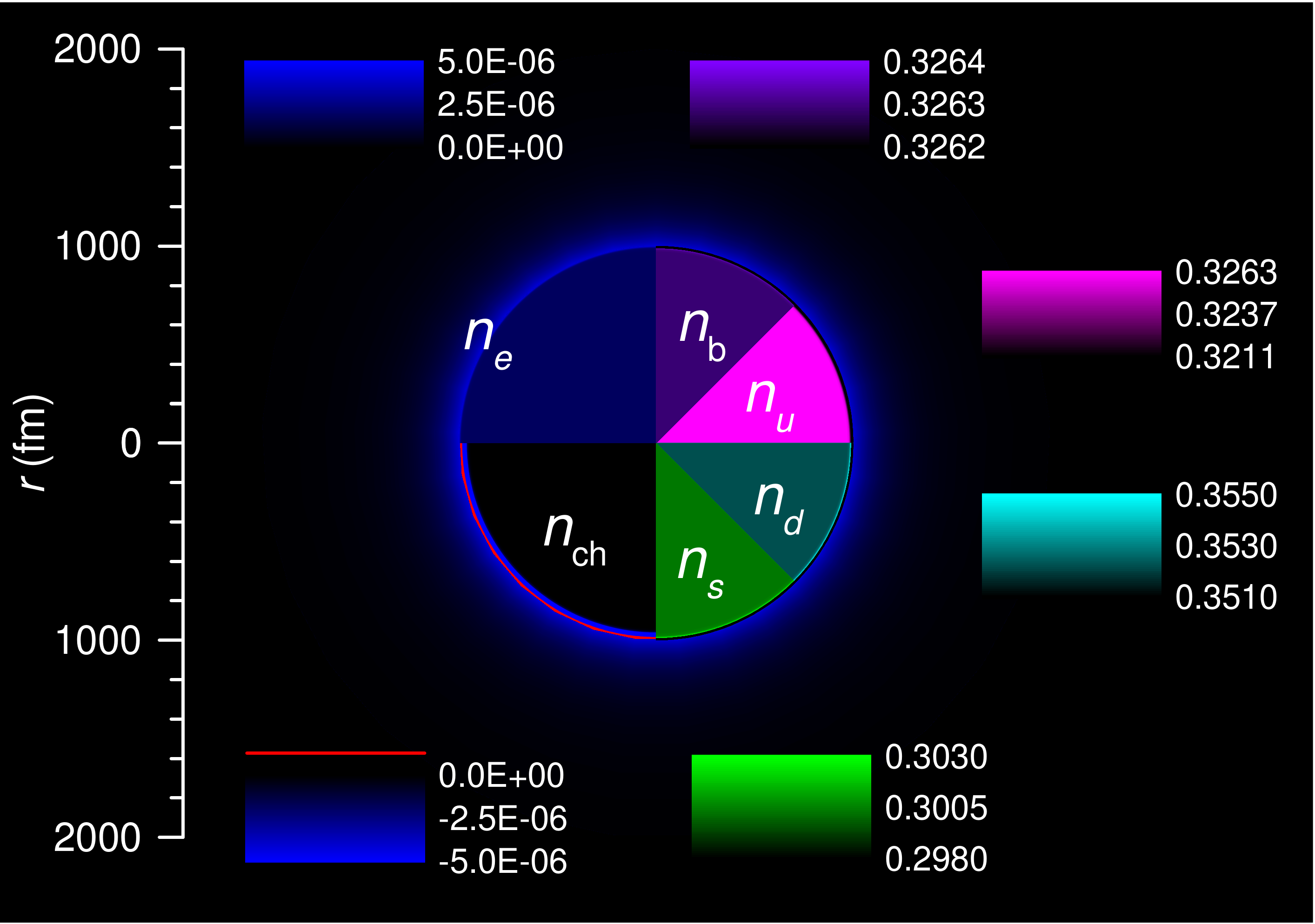}
\caption{\label{Fig:Slet3MRE}(Color online) The density profiles (in fm${}^{-3}$) of an SQM sphere with the core radius $R = 1000$ fm,
where the bag constant is taken as $B^{1/4} = 152$ MeV. Here $n_\mathrm{ch}=\sum_i q_i n_i$ corresponds to the local charge density, while
$n_\mathrm{b}=\sum_{q=u,d,s} n_q/3$ is the local baryon number density.}
\end{figure}

For a given core radius $R$, the structure of an SQM sphere is determined by solving the differential equation~(\ref{eq:pdis})
under the boundary conditions, i.e., $M_\mathrm{t}(0) = 0$, $Q(0)=Q(\infty)=0$, and Eq.~(\ref{eq:P_stable}). To illustrate our
results, we present a colored contour plot in Fig.~\ref{Fig:Slet3MRE} for an SQM sphere with the core radius $R=1000\ \mathrm{fm}$.
A rich charge profile on the surface is found. At the $r>R$ region, there exists an electron cloud which neutralizes the positively charged
core and expands by $\sim$$1\ \mathrm{\AA}$. Respectively, the SQM sphere contains $1.37\times10^{9}$ $u$-quarks,  $1.48\times10^{9}$
$d$-quarks, $1.26\times10^{9}$ $s$-quarks, and $2.74\times10^{5}$ electrons, which gives the total core charge $Q(R) = 2.66\times10^{5}$
and mass $M = 1.22\times10^{12}$ MeV. Due to the quark depletion on the core surface ($N_u^\mathrm{surf} = -3.34\times10^{4}$,
$N_d^\mathrm{surf} = -7.06\times10^{4}$, and $N_s^\mathrm{surf} = -1.03\times10^{6}$), there exists a surface charge
$Q^\mathrm{surf} = \sum_i q_i N_i^\mathrm{surf} = 3.43\times10^{5}$, while the corresponding mass modification is
$M^\mathrm{surf} = -2.15\times10^{8}$ MeV.

Note that the current masses of quarks and the electron mass are taken as $m_{u}=2.3$ MeV, $m_{d}=4.8$ MeV, $m_{s}=95$ MeV, and
$m_{e}=0.511$ MeV~\cite{Olive2014_CPC38-090001}, leaving only the bag constant $B$ undetermined. For SQM to stably exist
at zero external pressure, the bag constant should meet the requirement of the Witten-Bodmer hypothesis, which gives
$144.37< B^{1/4} < 159.26$ MeV. We take the three typical values $B^{1/4} = 145$, 152, and 159 MeV. It is worth mentioning
that if $B$ exceeds the upper bound, SQM is unstable and may only exist at the core of a compact star, i.e., hybrid
star~\cite{Weber2005_PPNP54-193, Wang2013_CSB58-3731}.

\begin{figure}
\includegraphics[width=\linewidth]{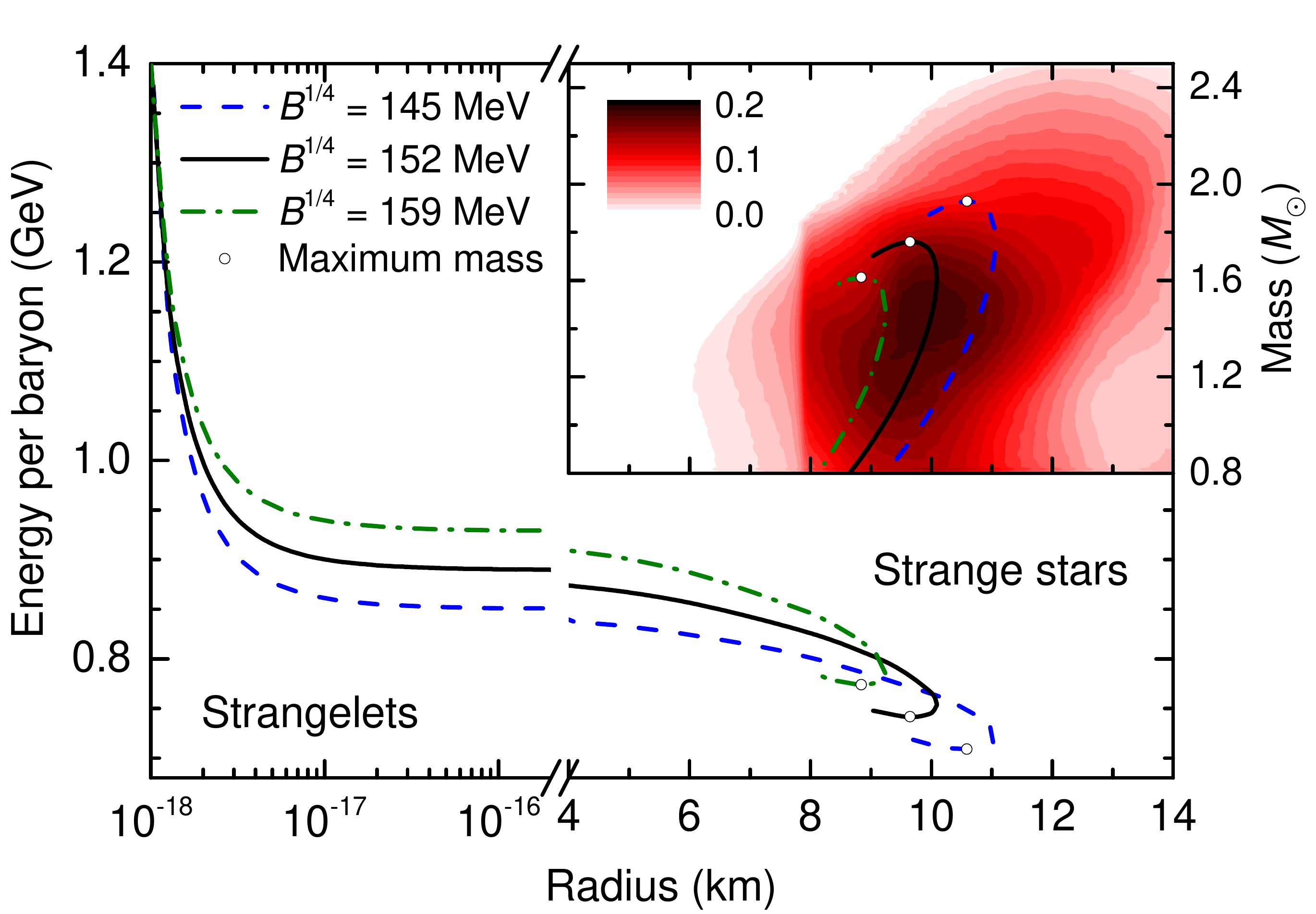}
\caption{\label{Fig:EPA}(Color online) The energy per baryon from slets to strange stars. In the inset,
the obtained $M$-$R$ relations are compared with the normalized $M$-$R$ probability distribution compiled from the observed values
of eight pulsars~\cite[Fig.~10]{Lattimer2012_ARNPS62-485}. }
\end{figure}

A full calculation from slets to strange stars is carried out. In Fig.~\ref{Fig:EPA}, the energy per baryon is given, which decreases
monotonously with increasing radius or baryon number. When the energy per baryon reaches 930 MeV, the minimum baryon numbers for absolutely
stable slets are determined, i.e., $A_\mathrm{min}= 24$, 80, and 394849, which increases dramatically as $B$ approaches to its upper limit.
In the region with $200\ \mathrm{fm} \lesssim R \lesssim 1\ \mathrm{km}$, the variation of the energy per baryon is infinitesimal, and the radius
is related to the baryon number by $R=r_0 A^{1/3}$ with $r_0=0.944$, 0.901, and 0.862 fm. When $R\gtrsim 1\ \mathrm{km}$, gravity starts to
reduce the energy per baryon and a minimum value is obtained, corresponding to the maximum mass and baryon number of strange stars.
The obtained $M$-$R$ relations, as shown in the inset of Fig.~\ref{Fig:EPA}, are in good agreement with the $M$-$R$ probability distribution
obtained by averaging the observed values of eight pulsars~\cite[Fig.~10]{Lattimer2012_ARNPS62-485}.

\begin{figure}
\includegraphics[width=\linewidth]{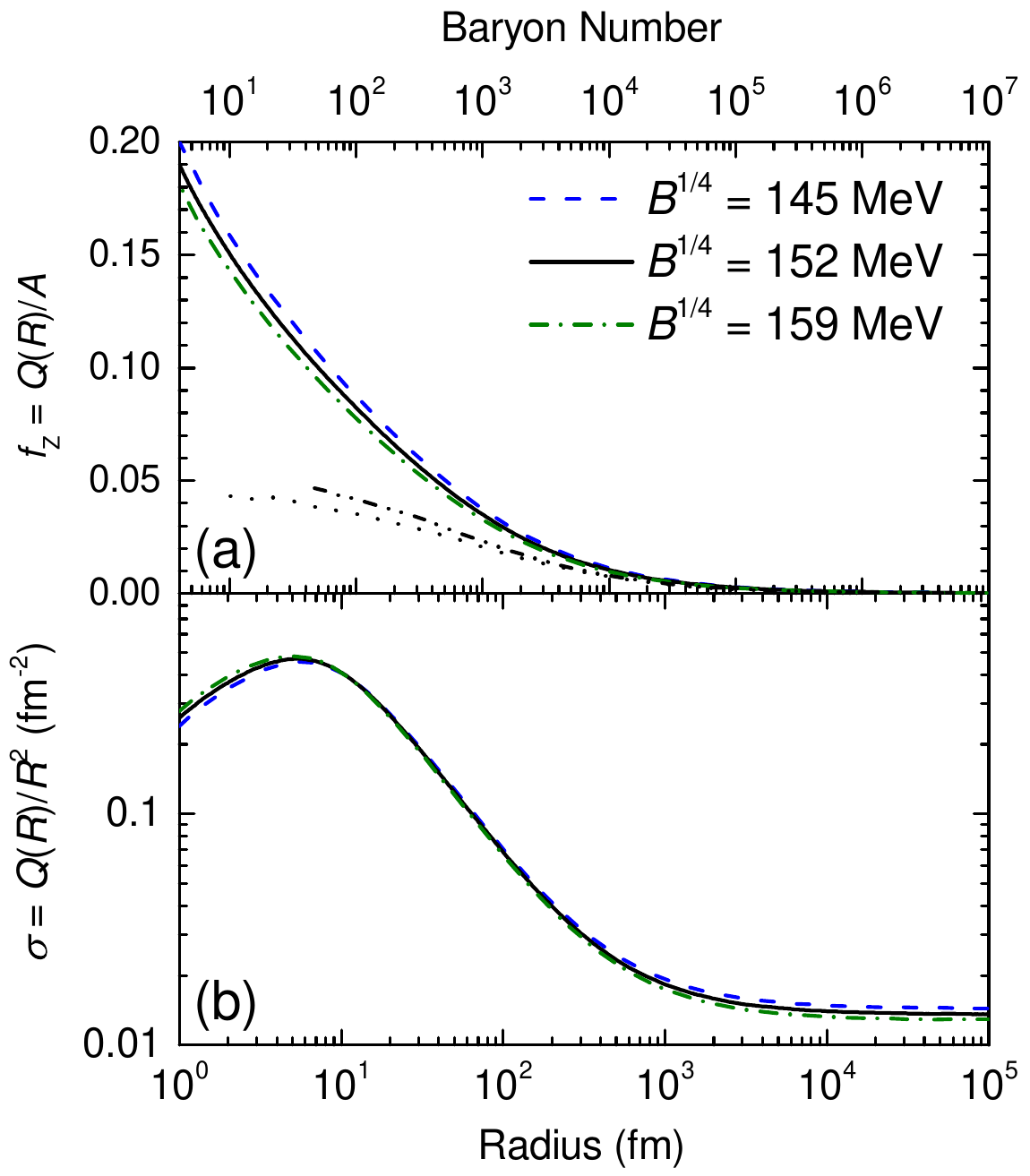}
\caption{\label{Fig:Charge}(Color online) (a): The charge-to-mass ratio of the SQM core of slets, which is compared with previous
findings indicated by the dotted line~\cite[Fig.~4]{Berger1987_PRC35-213} and dash-dot-dotted line~\cite[Fig.~2]{Heiselberg1993_PRD48-1418};
(b): The surface charge density of the SQM core as functions of radius.}
\end{figure}

The charge-to-mass ratio in Fig.~\ref{Fig:Charge}(a) is defined as $f_Z \equiv Q(R)/A$ ignoring the electrons surrounding the SQM core.
The obtained values are larger than previous results~\cite{Berger1987_PRC35-213, Heiselberg1993_PRD48-1418}. For smaller $B$, the SQM core
carries slightly more charge. Note that when $R\lesssim 13\ \mathrm{fm}$, SQM within slets is positively charged. However, for $R\gtrsim 17\
\mathrm{fm}$, as shown in Fig.~\ref{Fig:Slet3MRE}, the SQM carries negative charge to compensate the positive surface charge. Then an
electric potential well for negatively charged particles is formed due to the quark depletion on the quark-vacuum interface. These particles
may be trapped in the potential well and give a distinct photon spectrum when excited, which have significant implications for the experimental
searches of SQM. As indicated in Fig.~\ref{Fig:Charge}(b), when $R\gtrsim 10^{5}$ fm, charges are mostly located on the core surface and can
be described by a constant surface charge density $\sigma$ with $Q(R) = \sigma R^{2}$. It is found that $\sigma=0.0144$, 0.0135, and 0.0128
$\mathrm{fm}^{-2}$ for $B^{1/4} = 145$, 152, and 159 MeV, which are much larger than the upper bound $7\times10^{-5}\ \mathrm{fm}^{-2}$
considering the electron-positron pair creation~\cite{Madsen2008_PRL100-151102}.

\begin{figure}
\includegraphics[width=\linewidth]{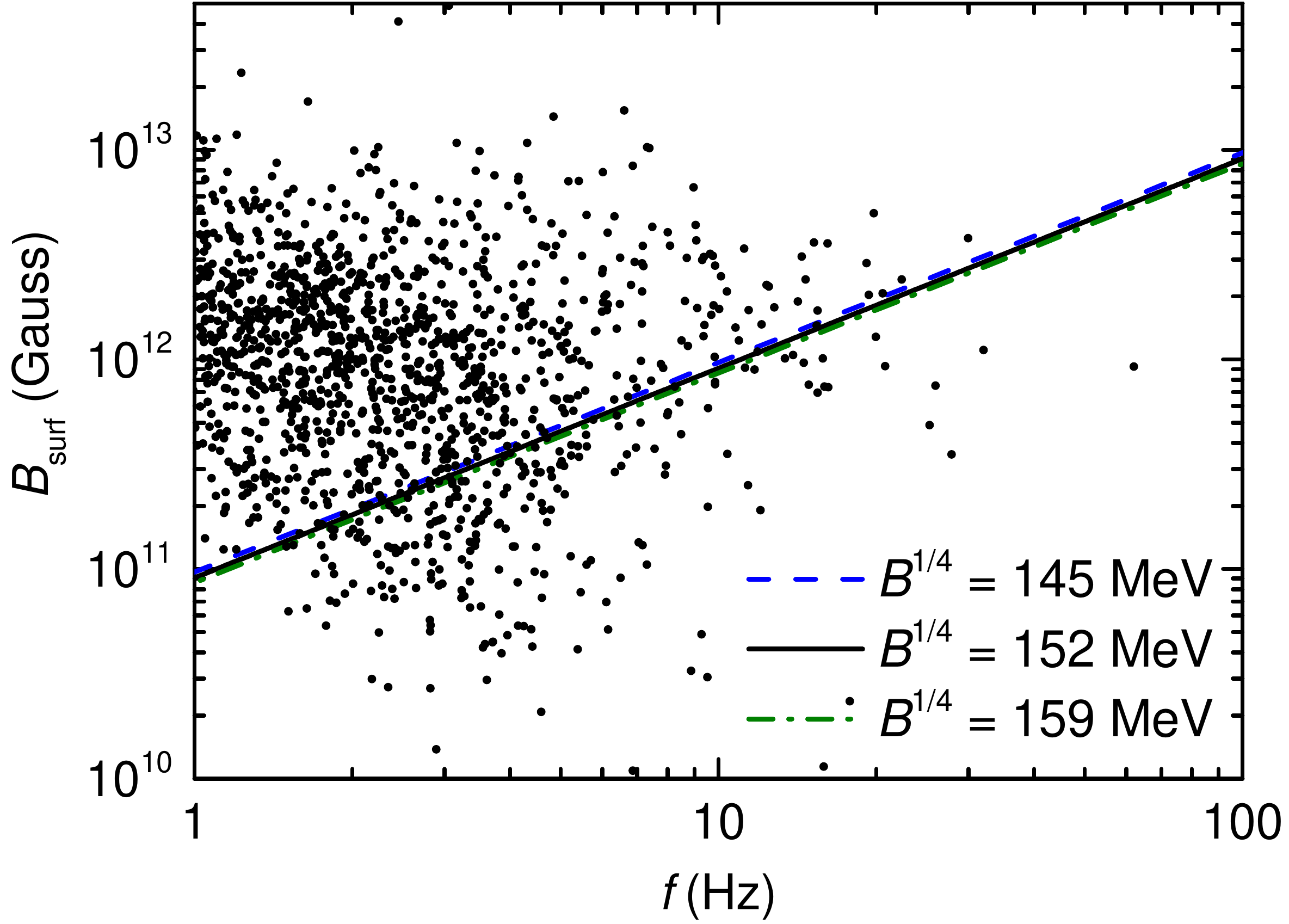}
\caption{\label{Fig:Magpole}(Color online) The magnetic field of rotating strange stars estimated with Eq.~(\ref{eq:Bsurf}).
The dots depict the inferred magnetic fields of pulsars obtained from ATNF Pulsar Catalogue~\cite{Manchester2005_AJ129-1993,
Manchester2015}, where the binary pulsars are excluded since their magnetic fields may be dampened by mass
accretion~\cite{Bhattacharya1991_PR203-1}. }
\end{figure}

When the SQM core rotates, a magnetic field may be generated. For strong enough field strength, the electrons are expected to be trapped
along the magnetic field lines. Then we simply assume the electron cloud stays still while the core rotates. It is straightforward to write
out the magnetic field at the pole area on the surface~\cite{Negreiros2010_PRD82-103010}:
\begin{equation}
 B_\mathrm{surf} = \frac{1}{3}u_0 \sigma R f. \label{eq:Bsurf}
\end{equation}
Here $u_0$ represents the vacuum permeability while $f$ is the rotational frequency. Then in Fig.~\ref{Fig:Magpole} the magnetic field of
a rotating strange star is obtained, with the field strength approaching to the typical value of pulsars, e.g., $\sim$$10^{12}$ Gauss at
$f=10$ Hz. It is found that the variation of the bag constant $B$ barely affects the field strength.

\section{\label{sec:sum}Summary}

In conclusion, we propose a unified description from strangelets to strange stars considering the gravity and electrostatic interactions on the
macroscopic scale while incorporating the strong and weak interactions locally. The quark-vacuum interface is treated with the multiple reflection
expansion method. The model parameter follows the Witten-Bodmer hypothesis and is confronted with the observational $M$-$R$ probability distributions
of pulsars. Then the properties of SQM systems with all possible baryon numbers are investigated. It is found that the energy per baryon decreases
monotonously for increasing baryon number, while the obtained charge-to-mass ratio of the SQM core is larger than previous predictions. On the core
surface, due to the quark depletion, an electric potential well is formed for negatively charged particles and may give some unique observables for
SQM detection. When $R\gtrsim 10^{5}$ fm, charges are mostly located on the core surface and a constant surface charge density is obtained.
Then for a rotational bare strange star, a magnetic field comparable to the typical strength of pulsars is generated.

\begin{acknowledgements}
We are grateful to Professors Lie-Wen Chen, Thomas Papenbrock, Michael Smith, and She-Sheng Xue for fruitful discussions.
This work was supported by National Natural Science Foundation of China (Grant Nos.~11135011, 11120101005, 11275248,
11475110,  11475115, 11575190 and 11525524), National Key Basic Research Program of China (Grant No.~2013CB834400),
and the Knowledge Innovation Project of the Chinese Academy of Sciences (Grant No. KJCX2-EW-N01). The computation of this work was supported by
the HPC Cluster of SKLTP/ITP-CAS and the Supercomputing Center, CNIC of CAS.
\end{acknowledgements}


\end{document}